\documentstyle[preprint,eqsecnum,aps,axodraw]{revtex}
\tighten
\begin{document}
\title{Compactification near and on the light front}
\author{{\bf A. Harindranath}$^{a,b}$, {\bf L$\!\!$'. Martinovi\v c}$^{a,c}$, and 
{\bf J. P. Vary}$^{a,d}$ \\
$^{a}$International Institute of Theoretical and Applied Physics, \\
Iowa State University, Ames, IA 50011, U.S.A. \\
$^{b}$Saha Institute of Nuclear Physics, 1/AF Bidhan Nagar, Calcutta, 700064,
India \\
$^{c}$ Institute of Physics, Slovak Academy of Sciences \\
D\'ubravsk\'a cesta 9, 842 28 Bratislava, Slovakia\\
$^{d}$Department of Physics and Astronomy, Iowa State University, Ames, IA 50011, U.S.A. \\}
\date{ December 9, 1999}
\maketitle
\begin{abstract}
We address problems associated with compactification near and
on the light front. In perturbative scalar field theory we illustrate and 
clarify the 
relationships among three approaches:
(1) quantization on a space-like surface close to a light front; (2)
infinite momentum frame calculations; and (3) 
quantization on the  light front. 
Our examples emphasize the difference between zero modes
in space-like quantization and those in light front quantization. In
particular, in perturbative calculations  
of scalar field theory
using discretized light cone quantization 
there are well-known new
$``$zero mode induced"
interaction terms. 
However, we show that they decouple in the continuum limit
and covariant answers are reproduced. Thus compactification of a light-like
surface is feasible and defines a consistent field theory.   
\end{abstract}
%%%%%%%%%%%%%%%%%%%%%%%%%%%%%%%%%%%%%%%%%%%%%%%%%%%%%%%%
\section{Introduction}
%%%%%%%%%%%%%%%%%%%%%%%%%%%%%%%%%%%%%%%%%%%%%%%%%%%%%%% 

Problems pertaining to compactification near and on the light front 
have been presented and explored 
recently in the context of perturbative scalar field 
theory\cite{pol}. In the formalism of compactification near the light front 
certain divergences were found in the one loop scattering amplitude in scalar
field theory {\it at finite box length} as one tried to approach the 
light front. These divergences were presumed to be caused by the
longitudinal zero momentum modes in the light front theory. 

Zero modes on the light front have a long history\cite{my,franke,lmr}. 
For  certain field
theories,
they are invoked
to account for the non-trivial vacuum structure. 
In order to isolate the zero mode, one puts the system in a longitudinal box
and imposes boundary conditions. 
This procedure is popularly known as Discretized Light Cone 
Quantization\cite{dlcq}
(DLCQ).
In DLCQ scalar field theory with periodic boundary conditions, 
the zero modes are
constrained and they have  to be determined in terms of the non-zero
modes by solving a nonlinear operator equation. Thus the zero mode in 
scalar light front theory is quite different from the 
zero mode in equal time theory where it
is a dynamical mode just as any non-zero mode. It is important to keep this
distinction in mind. 

Since zero modes pose a major challenge
in the nonperturbative context,
attempts have been made to perform the quantization on a space like 
surface\cite{nlf}
close to the light front (a parameter $\eta$ characterizes the 
$``$closeness").
By taking $ \eta \rightarrow 0$ one 
is supposed to reach the light front surface. However,
this limiting procedure need not be smooth since a light front surface cannot
be reached from a space-like surface by a {\it finite} Lorentz
transformation. On the other hand, S-matrix elements should be independent of
$\eta$ for {\it any} value of $\eta$ since this parameterization simply
labels
different space-like surfaces.  Thus any $\eta$ dependence in an S-matrix
element signals breakdown of Lorentz invariance as in the results of Ref.
\cite{pol}.

Let us recall the major differences between the discretized versions of 
near light front theory and
light front theory. 
We shall restrict the longitudinal coordinate $x^-$ 
($x^{\pm}=x^0\pm x^3$) to a 
finite interval while keeping two transverse coordinates unbounded.
To avoid confusion, we shall denote the light front box
length by $L$ and the near light front box length by $L_{et}$.
In order to check Lorentz invariance one has to perform the continuum limit
of DLCQ. Let us consider the mass operator $M^2=P^+ P^- - (P^\perp)^2$, 
 where $P^+, P^-$ are the light-front momentum and energy operators and 
$P^\perp \equiv (P^1,P^2)$. 
In DLCQ one can introduce 
$P^+ = { 2 \pi \over L}K$ and $ P^- = {  L \over 2 \pi} 
H$. The semi positive definite operator 
$K$, the
harmonic resolution, is dimensionless momentum and $H$, the Hamiltonian, has
the dimension of $M^2$.
In DLCQ the mass operator is given by $M^2 = KH - (P^\perp)^2$.
The box length $L$ has disappeared from the
operator. Eigenvalues of K represent the total momentum of the system. 
The continuum limit is given by $K \rightarrow \infty$. 
This is to be 
contrasted with the near light front discretization where the box length
does not disappear from the mass operator. Also the momentum operator is not
semi positive definite. Nevertheless for the ease of comparisons, let us
denote the total dimensionless momentum in the near 
light front
case by $K$. The longitudinal momentum $P$ in this case can take both
positive and negative values and we can put only $\mid P \mid = {2 \pi \over
L_{et}}K$. 

The infinite momentum frame\cite{ff,wein} is a concept that allows one to
simulate perturbative light front theory calculations in an equal time
framework by taking the external total longitudinal momentum to
infinity. In scalar field theory the equivalence holds even beyond tree
level (except for vacuum diagrams). 
%({\bf what happens to vacuum diagrams
%arising from zero mode induced terms in DLCQ as $[L_{lf}] L \rightarrow \infty$
%?} -- {\sl there are no such diagrams in scalar theory, because ZM induced 
%terms have no aaaa, $a^\dagger a^\dagger a^\dagger a^\dagger$ structures}). 
For theories involving fermions the equivalence clearly breaks down
beyond tree level\cite{dly,hp}. In scalar field theory, in the discretized
version, one can ask whether one can simulate DLCQ perturbation theory 
by considering the infinite momentum frame starting from the equal time
formulation. 
Obviously this cannot be achieved by taking $K$ very
large since that should correspond to the 
continuum limit of DLCQ. One choice is
to take $L_{et} \rightarrow 0$ since this can simulate infinite momentum for
non-zero modes. Then one can ask the question whether $L_{et}$ drops out of 
scattering amplitudes and if  they in turn approach DLCQ scattering 
amplitudes. 
Of course by taking $L_{et} \rightarrow 0$ we have moved as far away from
the continuum limit as possible and if we find Lorentz 
non-invariant answers we should not be 
surprised.  
Another choice is to discretize the near light front theory, let $ \eta 
\rightarrow 0$ and see whether $L_{et}$ dependence drops out 
(characteristic of the
DLCQ formalism).  At finite $\eta$, $ L \rightarrow \infty$ readily
reproduces covariant answers, but at finite $L$, $ \eta \rightarrow 0$
produces divergent answers. From this one cannot 
conclude anything about DLCQ since
Lorentz invariance is broken. Note 
that in the discretized 
near light front formulation, where modes are specified by   
integers $n$, the expression ${n \over \eta}$, encountered in 
\cite{pol}, presents for the zero mode ($n=0$) a
${0 \over 0}$ problem for $ \eta \rightarrow 0$ which means that the limit
is undefined.

To the best of our knowledge, various issues that are raised above  
have not been resolved in a clear and satisfactory manner.
Towards this goal,    
we perform and compare perturbative 
calculations for scalar field theory in the continuum and 
discretized versions of three
formulations, namely, light front quantization, infinite momentum limit of 
equal time quantization and space-like 
quantization parameterized by $\eta$.             
As examples we consider the self-energy diagram in $\phi^3$ theory and the 
scattering diagram in $\phi^4$ theory.

The plan of this paper is as follows.  In Sec. II we discuss the results in the 
continuum light front
theory. In Sec. III the corresponding DLCQ results are presented. 
In Sec. IV  the results in space like quantization characterized by $\eta$
are given in the continuum and the discretized versions. Corresponding
results in the infinite momentum limit of equal time theory are given in
Sec. V. Sec. VI contains our summary and conclusions. Since it is unfamiliar to
most readers, a brief introduction to 
the space-like quantization parameterized by $\eta$ 
is given in  Appendix A.   
%%%%%%%%%%%%%%%%%%%%%%%%%%%%%%%%%%%%%%%%%%%%%%%%%%%%%%%%%%%%
\section{Light front perturbation theory -- continuum formulation}
%%%%%%%%%%%%%%%%%%%%%%%%%%%%%%%%%%%%%%%%%%%%%%%%%%%%%%%%%%%%%
In this section we compare results of the light front perturbation theory 
with those of the covariant perturbation theory, 
both in continuum formulation. 
We consider the self-energy
diagram in ${\lambda \over 3!} \phi^3$ theory and the scattering
diagram  in ${\lambda \over 4!} \phi^4$ theory.
%%%%%%%%%%%%%%%%%%%%%%%%%%%%%%%%%%%%%%%%%%%%%%%%%%%%%%%%%%%%%%%%%%%%%%
\subsection{Self-energy  in ${\lambda \over 3!} \phi^3$ theory}
%%%%%%%%%%%%%%%%%%%%%%%%%%%%%%%%%%%%%%%%%%%%%%%%%%%%%%%%%%%%%%%%%%%%%%%
Consider the  one loop self-energy diagram in $ \phi^3$ theory.
Note that in this case
there is only one time ordered diagram (Fig. 1a) in the light front case.
Using the rules of light front old fashioned perturbation theory\cite{ks} 
we have
\begin{eqnarray}
\Sigma(p^2) = { 1 \over 2} \lambda^2 \int_0^{p^+} {dq^+ d^2 q^\perp \over 2 (2
\pi)^3} ~{ 1 \over q^+(p^+ - q^+)}~ 
{ 1 \over p^- - {(q^\perp)^2 + m^2 \over q^+} - {(p^\perp-q^\perp)^2 +m^2
\over p^+ - q^+}+ i \epsilon}. \label{3lfptc}
\end{eqnarray}
The factor $1/2$ is a symmetry factor.
Introducing $ y = q^+/p^+$, we get
\begin{eqnarray}
\Sigma(p^2) = { 1 \over 2} {\lambda^2 \over 2 (2 \pi)^3}\int_0^1 dy d^2 q^\perp 
{ 1 \over y(1-y)p^2 - (q^\perp)^2 -m^2 + i \epsilon}
\label{3lfpt}
\end{eqnarray}
with $ p^2 = p^+p^- - (p^\perp)^2$.  

Note that the integrand is
nonsingular at $y=0$. 

Next we derive this result starting from the Feynman diagram.
The corresponding amplitude (Fig. 2) is
\begin{eqnarray}
-i \Sigma(p^2)= {1 \over 2} {(-i\lambda)^2 \over(2 \pi)^4} \int d^4k~ 
{ i \over k^2
-m^2 + i \epsilon} ~{ i \over (p+k)^2 -m^2 + i \epsilon}.
\end{eqnarray}
Using $d^4k= { 1 \over 2} dk^+ dk^-
d^2 k^\perp$, we have
\begin{eqnarray}
\Sigma(p^2) &&= -{ i \over 2} {(-i\lambda)^2 \over (2 \pi)^4} { 1 \over 2}
\int_{-\infty}^{+ \infty} dk^+ \int d^2 k^\perp \int_{- \infty}^{+ \infty}
dk^- { 1 \over k^+(p^++k^+)}\nonumber \\
&&~~~ { 1 \over k^- - {(k^\perp)^2 + m^2 \over k^+} +
i { \epsilon \over k^+}} { 1 \over p^-+k^- - {(p^\perp + k^\perp)^2 + m^2
\over p^+ + k^+} + i {\epsilon \over p^+ + k^+}}.
\end{eqnarray}
Let us now perform the $k^-$ integration. 
Let $ p^+ > 0$. For $k^+ > 0$, $p^+ + k^+ >0$, both poles are in the lower half
of the complex $k^-$ plane. We can close the contour in the upper half plane
and the integral is zero.
For $k^+ < 0$, if $ p^+ < -k^+$, $ p^+ + k^+ <0$, both poles are in the
upper half plane. We can close the contour in the lower half plane and the
integral again is zero. For $p^+>0$, we get a non-vanishing contribution
when $ k^+ <0$ and $k^+ > -p^+$. Then, closing the contour in the upper
half plane, we get
\begin{eqnarray}
\Sigma(p^2) = { 1 \over 2} {(-i\lambda)^2 \over 2 (2 \pi)^3}\int_{-p^+}^0 {dk^+
d^2 k^\perp \over k^+(p^++k^+)}~ 
{ 1 \over p^- + {(k^\perp)^2 + m^2 \over k^+} - {(p^\perp -k^\perp)^2 + m^2
\over p^++k^+} - i {\epsilon \over k^+} + i {\epsilon \over p^++k^+}}
\end{eqnarray}
or
\begin{eqnarray}
\Sigma(p^2) = { 1 \over 2} \lambda^2 \int_0^{p^+} {dq^+ d^2 q^\perp \over 2 (2
\pi)^3} ~{ 1 \over q^+(p^+ - q^+)}~ 
{ 1 \over p^- - {(q^\perp)^2 + m^2 \over q^+} - {(p^\perp-q^\perp)^2 +m^2
\over p^+ - q^+}+ i \epsilon}.
\end{eqnarray}
We recover the expression (Eq. (\ref{3lfptc})) from old 
fashioned perturbation theory with energy
denominator and integration over three momentum. 
%%%%%%%%%%%%%%%%%%%%%%%%%%%%%%%%%%%%%%%%%%%%%%%%%%%%%%%%%%%%%%%%%%%%
\subsection{One loop scattering in ${\lambda \over 4!} \phi^4$ theory}
%%%%%%%%%%%%%%%%%%%%%%%%%%%%%%%%%%%%%%%%%%%%%%%%%%%%%%%%%%%%%%%%%%%%%
Next consider the scattering amplitude at one loop level in $ \phi^4$ theory.
$p_1,p_2$ are the initial momenta and $p_3,p_4$ are the final momenta. 
 Let us denote $s= (p_1+p_2)^2$ and $ t = (p_1-p_3)^2$. 
In the light front perturbation theory, we have to consider two 
cases separately. 

1) $p_1^+ > p_3^+$. 

The scattering amplitude (Fig. 3a) in this case is
\begin{eqnarray}
M_{fi} && = {1 \over 2}{ \lambda^2 \over 2 (2 \pi)^3} \int_0^{p_1^+ - p_3^+} 
dq_1^+ 
\int d^2 q_1^\perp ~{ 1 \over q_1^+} ~{ 1 \over p_1^+ - p_3^+ - q_1^+}
\nonumber \\
&&~~~~~~~~~{ 1 \over p_1^- + p_2^- - p_3^- -p_2^- - q_1^- - (p_1-p_3-q_1)^-} 
\nonumber \\
&& = { 1 \over 2}{ \lambda^2 \over 2 (2 \pi)^3} \int_0^{p_1^+ - p_3^+} dq_1^+ 
\int d^2 q_1^\perp ~{ 1 \over q_1^+} ~{ 1 \over p_1^+ - p_3^+ - q_1^+}
~{ 1 \over p_1^- - p_3^- -q_1^- - (p_1-p_3-q_1^-} \nonumber \\
&& = \theta(p_1^+ - p_3^+) { 1 \over 2}{ \lambda^2 \over 2 (2 \pi)^3} \int_0^1 dy \int
d^2 q^\perp { 1 \over y (1-y) t - (q^\perp)^2 -m^2 + i \epsilon}.
\label{lff1}
\end{eqnarray}
We have used the notation $(p-q)^-={(p^\perp - q^\perp)^2 + m^2 \over p^+ -
q^+}$.

2) $p_1^+ < p_3^+$. 

The scattering amplitude (Fig. 3b) in this case is  
\begin{eqnarray}
M_{fi} && ={1 \over 2} { \lambda^2 \over 2 (2 \pi)^3} \int_0^{p_3^+ - p_1^+} 
dq_2^+ 
\int d^2 q_2^\perp ~{ 1 \over q_2^+} ~{ 1 \over p_3^+ - p_1^+ - q_2^+}
\nonumber \\
&&~~~~~~~~{ 1 \over p_1^- + p_2^- - p_1^- -p_4^- - q_2^- - (p_1-p_3-q_2)^-} \nonumber
\\
&& = { 1 \over 2}
{ \lambda^2 \over 2 (2 \pi)^3} \int_0^{p_3^+ - p_1^+} dq_2^+ 
\int d^2 q_2^\perp ~{ 1 \over q_2^+} ~{ 1 \over p_3^+ - p_1^+ - q_2^+}
~{ 1 \over p_3^- - p_1^ - q_2^- - (p_3-p_1-q_2)^-} \nonumber \\
&& = \theta(p_3^+ - p_1^+) { 1 \over 2}
{ \lambda^2 \over 2 (2 \pi)^3} \int_0^1 dy \int
d^2 q^\perp ~{ 1 \over y (1-y) t - (q^\perp)^2 -m^2 + i \epsilon}.
\label{lff2}
\end{eqnarray}
We have used overall energy conservation $ p_1^- + p_2^- = p_3^- + p_4^-$ and
hence $ p_2^- - p_4^- = p_3^- - p_1^-$. 
Adding the two contributions we get 
\begin{eqnarray}
M_{fi} =  { 1 \over 2}{ \lambda^2 \over 2 (2 \pi)^3} \int_0^1 dy \int
d^2 q^\perp { 1 \over y (1-y) t - (q^\perp)^2 -m^2 + i \epsilon}.
\label{lft}
\end{eqnarray}
Note that the integrand is nonsingular at $y=0$. 

Starting from the Feynman amplitude,
the scattering amplitude (Fig. 4) is 
\begin{eqnarray}
-iM_{fi} &&={1 \over 2} {(-i \lambda)^2 \over (2 \pi)^4} \int d^4q ~{ i \over q^2 -m^2+i
\epsilon} ~{ i \over (p_1-p_3-q)^2 -m^2 + i \epsilon} \nonumber \\
M_{fi}&& ={1 \over 2} i{ \lambda^2 \over (2 \pi)^4} { 1 \over 2} \int dq^+ d^2 q^\perp
dq^-
{ 1 \over q^+ (p_1^+ - p_3^+ -q^+)} \times \nonumber \\
&&~~~ { 1 \over q^- - {(q^\perp)^2 + m^2 \over q^+} + i { \epsilon \over
q^+}} ~{ 1 \over p_1^- - p_3^- - q^- - {(p_1^\perp -p_3^\perp - q^\perp)^2
\over p_1^+ - p_3^+ - q^+} + i { \epsilon \over p_1^+ - p_3^+ - q^+}}.  
\end{eqnarray}
Now we have to distinguish two cases separately.

1) $p_1^+ - p_3^+ > 0$.

Non-vanishing contribution can occur only when $ q^+ >0$ and $ p_1^+ - p_3^+-
q^+ >0$. Then poles appear in both upper and lower half planes in the complex
$k^-$ plane. Closing the contour in the lower half plane, we get,
\begin{eqnarray}
M_{fi} &&={1 \over 2} i{\lambda^2 \over (2 \pi)^4} { 1 \over 2} (-2 \pi i) \int_0^{p_1^+ -
p_3^+} dq^+ \int d^2 q^\perp { 1 \over q^+(p_1^+ - p_3^+ - q^+)} \nonumber
\\
&&~~{ 1 \over p_1^-- p_3^- -{(q^\perp)^2 + m^2 \over q^+} - {(p_1^\perp -
p_3^\perp - q^\perp)^2 + m^2 \over p_1^+ - p_3^+ - q^+} + i \epsilon} .
\end{eqnarray}   

2) $ p_1^+ - p_3^+ <0$.

Non vanishing contribution can occur only when $q^+ <0$ and $ p_1^+ - p_3^+ -
q^+ <0$. Closing the contour in the upper half plane, we get
\begin{eqnarray}
M_{fi} &&= {1 \over 2}i{ \lambda^2 \over (2 \pi)^4} { 1\over 2} ( 2 \pi i) \int_{p_1^+ -
p_3^+}^{0} dq^+ \int d^2 q^\perp { 1 \over q^+ (p_1^+ - p_3^+ - q^+)}
\nonumber \\
&&~~ { 1 \over p_1^- - p_3^- - { (q^\perp)^2 + m^2 \over q^+} - 
{ (p_1^\perp - p_3^\perp -q^\perp)^2 + m^2 \over p_1^+ - p_3^+ - q^+} + i
\epsilon}.
\end{eqnarray}
Thus, in this case, we reproduce the two time ordered diagrams in old  fashioned
perturbation theory (Fig. 3).

After a change of variable in the second contribution, the two contributions
can be combined to yield
\begin{eqnarray}
M_{fi} = {1 \over 2}{ \lambda^2 \over 2(2 \pi)^3} \int_0^1 dy
\int d^2 q^\perp  { 1 \over y(1-y)t - (q^\perp)^2  - m^2 +
i \epsilon}.
\end{eqnarray} 
%%%%%%%%%%%%%%%%%%%%%%%%%%%%%%%%%%%%%%%%%%%%%%%%%%%%%%%%%%%%%%%%%%%%%%%%%%%%
%%%%%%%%%%%%%%%%%%%%%%%%%%%%%%%%%%%%%%%%%%%%%%%%%%%%%%%%%%
\section{Light front perturbation theory -- discretized 
formulation} 
%%%%%%%%%%%%%%%%%%%%%%%%%%%%%%%%%%%%%%%%%%%%%%%%%%%%%%%%%%%
Light front quantization in a finite volume with periodic fields (DLCQ) has some  
conceptual advantages. First of all, it allows one to work explicitly with 
Fourier modes of quantum fields, carrying vanishing light front  momentum $p^+$ -- the 
zero modes (ZM). While in the case of gauge fields some ZM are dynamically 
independent, ZM of scalar fields are always dependent (constrained) variables, 
as follows from the structure of the equations of motion, containing 
$\partial_\mu\partial^\mu = 
4\partial_+\partial_- -\partial_\perp^2, \partial_\perp^2 \equiv \partial_i
\partial_i, i=1,2$. Due to periodic boundary conditions 
\footnote{We use finite interval for all three space coordinates  
in this section.} in $x^-$ and $x^\perp \equiv (x^1,x^2)$ ($-L \leq x^- \leq L, -L_\perp \leq 
x^\perp \leq L_\perp$), the full scalar field can be 
decomposed as $\phi(x) = \phi_0(x^+,x^\perp) 
 + \phi_n(x^+,\underline{x})$, where $\underline{x} \equiv (x^-,
x^\perp)$. The mode expansion for the normal-mode field $\phi_n(\underline{x})$ 
is
\begin{eqnarray}
\phi_n(\underline{x}) = {1 \over {\sqrt{V}}}\sum_{\underline{k}}{1 \over 
{\sqrt{k^+}}}
\left[a_{\underline{k}}e^{-i\underline{k}\underline{x}} + a^{\dagger}_
{\underline{k}}e^{i\underline{k}\underline{x}} \right].
\label{phiexp}
\end{eqnarray}
Here we have used the notation $\underline{k}\underline{x} \equiv {1\over 2}
k^+x^- - k^\perp x^\perp$ and $k^+={2\pi \over L}n, n=1,2,\dots N,\; k^\perp = 
{2\pi \over L_\perp}n^\perp, n^\perp = 0,\pm 1,\pm 2,\dots \pm N_\perp$. In 
the following, 
the integration over the 3-dimensional volume will be denoted by $\int_{V}d^3
\underline{x} \equiv {1 \over 2}\int_{-L}^L dx^-\int_{-L_\perp}^{L_\perp}
d^2x^\perp$. 
%%%%%%%%%%%%%%%%%%%%%%%%%%%%%%%%%%%%%%%%%%%%%%%%%%%%%%%%%%%
\subsection{$\phi^3$ theory}

The DLCQ Hamiltonian of the $\phi^3$ theory, obtained in the canonical 
way, is
\begin{eqnarray}
P^- = \int_{V}d^3\underline{x}\left[m^2\phi^2 + (\partial_\perp \phi)^2 + 
{\lambda \over 3}\phi^3 \right].
\label{Ham3}
\end{eqnarray} 
It contains ZM terms, which have to be expressed by means of the normal-mode  
field $\phi_n(\underline{x})$. To do so we need to obtain the lowest-order 
solution of the ZM constraint. The latter is simply the ZM projection of the 
equation of motion 
\begin{eqnarray}
(4\partial_+ \partial_- - \partial_{\perp}^2)\phi = -m^2\phi -{\lambda \over 2}
\phi^2
\end{eqnarray}
and reads
\begin{eqnarray}
(m^2 - \partial_{\perp}^2)\phi_0 = -{\lambda \over 2}\int_{-L}^{L}{dx^- \over 
{2L}}(\phi_0^2 + \phi_n^2). 
\label{constr}
\end{eqnarray}
It can be solved iteratively and to the lowest order in $\lambda$ one has 
\begin{eqnarray}
\phi_0 = -{\lambda \over 2}{1 \over {m^2 - \partial_{\perp}^2}} \int_{-L}^L 
{dx^- \over {2L}}\phi_n^2.
\label{lorder}
\end{eqnarray}
The symbolic inverse operator $(m^2 -\partial_{\perp}^2)^{-1}$ is defined in  
momentum representation by replacing $\partial_{\perp}^2$ by the minus square 
of the perpendicular momentum of the composite operator in the integrand. 
In the Fock representation, one finds
\begin{eqnarray}
\phi_0(x^\perp)=-{\lambda \over V}\sum_{\underline{k}_1,\underline{k}_2}
{{\delta_{k_1^+,
k_2^+}}\over{\sqrt{k_1^+ k_2^+}}}{{e^{-i(k_1^{\perp}-k_2^{\perp})x^{\perp}}}
\over{m^2 + (k_1^{\perp} - k_2^{\perp})^2}}a^{\dagger}_{\underline{k}_1} 
a_{\underline{k}_2} - {\lambda \over {2m^2}}{1 \over V}\sum_{\underline{k}_1}
{1 \over k_1^+},
\label{Fockconstr}
\end{eqnarray}
where the second term comes from the normal ordering. This term will be 
neglected henceforth because it generates divergent terms in the Hamiltonian, 
which are presumably a manifestation of the well known pathology of the 
$\lambda\phi^3$ theory (no lower bound of the energy). Indeed, in the case    
of the $\lambda\phi^4$ interaction, the constrained zero mode is expressed 
automatically as a normal-ordered product of creation and annihilation 
operators without a c-number piece.      

The interacting Hamiltonian $P^-_{int}$ contains a term, corresponding to the  
usual one of the continuum formulation, plus 
the ZM term, calculated to $O(\lambda^2)$:  
\begin{eqnarray}
P^-_{int} = P_{NM} + P_{ZM}^{-(2)},\;\;\; 
P_{NM} = {\lambda \over 3}\int_V d^3{\underline{x}}\phi_n^3, 
\label{Hamint3}
\end{eqnarray}
\begin{eqnarray}
P_{ZM}^{-(2)}= \int_V d^3{\underline{x}}\left[\phi_0 (m^2 - \partial_{\perp}^2) 
\phi_0 + {\lambda \over 3}(\phi_0\phi_n^2+\phi_n\phi_0\phi_n+ \phi_n^2\phi_0
)\right] 
\label{ZMHam3}
\end{eqnarray}   
with $\phi_0$ given by Eq.(\ref{lorder}). The symmetric operator ordering has 
been used in the last term.  
The $O(\lambda^2)$ self-energy amplitude, corresponding to the first term in 
(\ref{Hamint3}), can be calculated by the old fashioned perturbation theory 
formula 
\begin{eqnarray}
T_{fi} = \sum_{n}{{\langle \underline{p}^\prime \vert P_{NM}\vert n \rangle 
\langle n \vert P_{NM} \vert \underline{p} \rangle}\over{p^- - p^-_{n}}}, 
\label{PTformula}  
\end{eqnarray}
where $H_I$ denotes the interacting Hamiltonian, $\vert \underline{p} \rangle 
\equiv a^{\dagger}_{\underline{p}}\vert 0 \rangle$ and the summation runs over 
the two-particle intermediate states $\vert n \rangle \equiv 
2^{-{1 \over 2}}a^{\dagger}_{\underline{i}_1}a^{\dagger}_{\underline
{i}_2}\vert 0 \rangle$. After inserting the field expansion (\ref{phiexp}) and 
performing the operator commutations, we arrive at
\begin{eqnarray}
T_{fi}= {\delta_{\underline{p},\underline{p^\prime}}
\over{\sqrt{p^+Vp^{\prime +}V}}} {\lambda^2 \over 4}\sum_{\underline{q}}
{1 \over {q^+(p^+-q^+)}}{1 \over {{(p^\perp)^2+m^2}\over{p^+}} - {{(q^\perp)^2+
m^2}\over{q^+}} - {{(p^\perp-q^\perp)^2 + m^2}\over{p^+-q^+}}},
\label{DLCQ3}
\end{eqnarray}
where $q^+ < p^+= 2\pi K L^{-1}, \vert q^\perp \vert < 2\pi \Lambda_\perp 
L_\perp^{-1}$ and $K, \Lambda_\perp$ are integers. From this expression, the 
continuum answer for the self-energy $\Sigma(p^2)$ (\ref{3lfptc}) or  
(\ref{3lfpt}) can be extracted in the infinite volume limit 
$K,L,\Lambda_\perp,L_\perp \rightarrow \infty $ ($p^+$ kept fixed) with 
${1 \over V}\Sigma_
{\underline{q}} \rightarrow {1 \over {(2\pi)^3}}\int {dq^+ \over 
2} d^2q_{\perp}$, ${V \over 2} \delta_{\underline{p},\underline{k}} \rightarrow 
(2\pi)^3\delta(\underline{p} - \underline{q})$. We recall that 
$\Sigma$ corresponds to the invariant amplitude $M_{fi}$ which differs by   
$(2\pi)^3$ times the kinematical factor (first term in (\ref{DLCQ3})) from 
$T_{fi}$.  

The ZM Hamiltonian does not contribute in the continuum limit. Indeed,  
the first term in (\ref{ZMHam3}) to $O(\lambda^2)$ is  
\begin{eqnarray}
P^{-(2)}_{ZM_1} &&= {1 \over 2}{\lambda^2 \over V}\Big\{ \sum_{\underline
{k}_1\dots\underline{k}_4} {{\delta_{k_1^+,k_2^+}\delta_{k_3^+,k_4^+}}\over{
\sqrt{k_1^+k_2^+k_3^+k_4^+}}}\delta^2_{k^\perp_1 + k^\perp_3,k^\perp_2 + 
k^\perp_4} {{a^\dagger_{\underline{k}_1}a^\dagger_{\underline{k}_3}
a_{\underline{k}_2} a_{\underline{k}_4}} \over {m^2 + (k^\perp_1-k^\perp_2)^2}}
\nonumber \\
&&~~~~+ \sum_{\underline{k}_1, k^{\perp}_2}{1 \over {k_1^{+2}}} 
{{a^{\dagger}_{\underline{k}_1} a_{\underline{k}_1}}\over{m^2 + 
(k^{\perp}_1 - k^{\perp}_2)^2}} \Big\}. 
\end{eqnarray}
The second term of (\ref{ZMHam3}) has the same structure with the individual  
coefficients $-1$ and $-{2 \over 3}$ instead of the overall 
$1 \over 2$ and thus the full $O(\lambda^2)$ ZM Hamiltonian is equal to   
\begin{eqnarray}
P^{-(2)}_{ZM}&& = -{1 \over 2}{\lambda^2 \over V}\Big\{
\sum_{\underline
{k}_1\dots\underline{k}_4} {{\delta_{k_1^+,k_2^+}\delta_{k_3^+,k_4^+}}\over{
\sqrt{k_1^+k_2^+k_3^+k_4^+}}}\delta^2_{k^\perp_1 + k^\perp_3,k^\perp_2 + 
k^\perp_4} {{a^\dagger_{\underline{k}_1}a^\dagger_{\underline{k}_3}
a_{\underline{k}_2} a_{\underline{k}_4}} \over {m^2 + (k^\perp_1-k^\perp_2)^2}}
\nonumber \\
&&~~~~+ {1 \over 3}\sum_{\underline{k},q^{\perp}}
{1 \over {k^{+2}}}{{a^{\dagger}_{\underline{k}}a_{\underline{k}}}\over {m^2 
+ (k^{{\perp}} - q^{\perp})^2}}\Big\}. 
\end{eqnarray}
Its contribution to the boson self-energy in the first order perturbation 
theory is  
\begin{eqnarray}
\tilde{T}_{fi} = -{1 \over 6}{\delta_{\underline{p^{\prime}},\underline{p}} 
\over {\sqrt{p^+ V p^{\prime +} V}}}
{\lambda^2 \over p^+}
\sum_{q^\perp}{1 \over {m^2 + (p^{\perp}-q^{\perp})^2}}.
\label{ZMamp3}
\end{eqnarray}
The corresponding $M$-amplitude vanishes in the continuum limit  
due to the extra $L^{-1}$ factor (a similar result in the case of 
${\lambda \over {4!}}
\phi^4(1+1)$ 
has been obtained in Ref. \cite{Heinzl}:
\begin{eqnarray}
\tilde{\Sigma}(p^+,p^\perp) = -{1 \over 6}{\lambda^2 \over {(2\pi)^2}}
{1\over L}{1 
\over {p^+}} \int d^2 q^{\perp}{1 \over {m^2 + (p^{\perp}-q^{\perp})^2}}.
\label{M3}
\end{eqnarray}
%volume integral times $L_{\perp}^2 \over {\pi^2}$, we find that $\tilde
%{\Sigma}$ logarithmically diverges, similarly as the full continuum self-energy 
%(\ref{3lfpt}). However, $\tilde{\Sigma}$ is additionally  
%suppressed by $K$ and thus it vanishes in the continuum limit $K \rightarrow 
In this way, DLCQ calculation yields the correct covariant result   
for the one-loop self-energy in $\lambda\phi^3$ 
theory in the infinite-volume limit.   
%%%%%%%%%%%%%%%%%%%%%%%%%%%%%%%%%%%%%%%%%%%%%%%%%%%%%%%%%%%

\subsection{$ \phi^4$ theory}

In order to calculate the one-loop scattering amplitude in DLCQ perturbation 
theory for $(4!)^{-1}\phi^4$ (3+1) model, we again need to derive the light
front 
Hamiltonian 
with $O(\lambda^2)$ ZM effective interactions. Following the same steps as in 
the previous subsection with $(3!)^{-1}\lambda \phi^3$ interaction replaced by  
$(4!)^{-1}\lambda \phi^4$, we find
\begin{eqnarray}
P^-_{int} = {2\lambda \over 4!}\int_{V}d^3{\underline{x}}\;\phi^4_n(\underline{x})
\;\;+ \;\;P^{-(2)}_{ZM},
\end{eqnarray}
where the second-order ZM Hamiltonian is
\begin{eqnarray}
P^{-(2)}_{ZM} = \int_{V}d^3{\underline{x}}\left[\phi_0(m^2 - \partial_{\perp}^2
)\phi_0 + {2\lambda \over{4!}}\; 4\phi_0\phi_n^3\right]. 
\end{eqnarray}
In the last term, the symmetric operator ordering between the lowest-order 
solution of the ZM constraint  
\begin{eqnarray}
\phi_0 = -{\lambda \over 3!}{1 \over {m^2 - \partial_{\perp}^2}} \int_{-L}^L 
{dx^-\over{2L}}\phi_n^3
\label{lorder4}
\end{eqnarray}
and $\phi_n^3$ is assumed. In the Fock representation, one obtains
\begin{eqnarray}
\phi_0(x^\perp) && = -{\lambda \over 2}{1 \over V^{3\over 2}} \sum_{\underline{k}
_1,\underline{k}_2,\underline{k}_3}{1 \over \sqrt{k^+_1 k^+_2 k^+_3}}{\delta_{
k^+_1,k^+_2 + k^+_3} \over {m^2 + (k^\perp_1 - k^\perp_2 - k^\perp_3)^2}}
\nonumber \\
&&~~~~\left[a^\dagger_{\underline{k}_3} a^\dagger_{\underline{k}_2}a_{\underline{k}_1}
e^{-i(k^\perp_3 + k^\perp_2 - k^\perp_1)x^\perp} + h.c. \right].
\label{Fockconstr4}
\end{eqnarray}
Using the formula (\ref{PTformula}) with $\vert \underline{p} \rangle  
\rightarrow \vert \underline{p}_1,\underline{p}_2 \rangle $, $\vert \underline
{p}^\prime \rangle \rightarrow \vert \underline{p}_3,\underline{p}_4 \rangle $  
and with four-particle intermediate states, one finds after a lot of algebra for 
the second-order NM scattering amplitude the expression 
\begin{eqnarray}
T_{fi} &&= {\delta_{\underline{p}_4+\underline{p}_3,\underline{p}_2+\underline{p}
_1}\over{\sqrt{p^+_4 V p^+_3 V p^+_2 V p^+_1 V}}}{\lambda^2 \over 4}\sum_
{\underline{q}}{1 \over{q^+(p^+_3-p^+_1-q^+)}}{1 \over{p^-_3 - p^-_1 - (p_3-
p_1-q)^-} } + (1 \leftrightarrow 3) 
\label{DLCQ4} \nonumber \\
\end{eqnarray}
The continuum-limit invariant scattering amplitude $M_{fi}$, extracted from 
(\ref{DLCQ4}), coincides  
with the covariant answer (\ref{lft}). It follows that for consistency the ZM 
contribution has to vanish in the continuum limit. That this is indeed the case 
can be checked in the 
first-order perturbation theory. In the Fock representation,  
part of the ZM Hamiltonian  
relevant for 
$2\rightarrow 2$ scattering, takes 
the form
\begin{eqnarray}
P^{-(2)}_{ZM_4} = -{\lambda^2 \over 4}{1 \over {V^2}}\sum_{q^\perp}\sum_{\underline
{k}_4,
\underline{k}_3,\underline{k}_2,\underline{k}_1}{{\delta_{\underline{k}_4 + 
\underline{k}_3,\underline{k}_2 + \underline{k}_1}}\over{\sqrt{k^+_4 k^+_3 
k^+_2 k^+_1}}}{1 \over{k^+_1 - k^+_2}}{{a^{\dagger}_{\underline{k}_3}a^{\dagger}
_{\underline{k}_1}a_{\underline{k}_4}a_{\underline{k}_2}}\over{m^2 + (q^\perp 
+k^{\perp}_2-k^{\perp}_1)^2}}.
\label{ZMH4}
\end{eqnarray}  
The corresponding scattering amplitude is
\begin{eqnarray}
\tilde{T}_{fi} = -{{\delta_{\underline{p}_4 + 
\underline{p}_3, 
\underline{p}_2 + \underline{p}_1}}\over{\sqrt{p^+_4 V p^+_3 V p^+_2 V p^+_1 V 
}}}{\lambda^2 \over 8}{1 
\over {p^+_3 - p^+_1}} \sum_{q^\perp}{1 \over{m^2 + (q^\perp + p^{\perp}_1 - 
p^{\perp}_3)^2}} + (1 \leftrightarrow 3)
\label{ZMamp4}
\end{eqnarray}
and the invariant amplitude $\tilde{M}$ indeed vanishes for $L \rightarrow 
\infty$:  
\begin{eqnarray}
\tilde{M}_{fi}(p^+_3-p^+_1,p^{\perp}_3-p^{\perp}_1)&& = -{\lambda^2 \over{ 8(2\pi)
^3}}{1 \over{p^+_3 - p^+_1}}{1 \over 
{L}} \int d^2 q^\perp{1 \over{m^2 + (q^\perp + p^{\perp}_1 - p^{\perp}_3)^2}} 
+ (1 \leftrightarrow 3).
\label{M4} \nonumber \\
\end{eqnarray}
%%%%%%%%%%%%%%%%%%%%%%%%%%%%%%%%%%%%%%%%%%%%%%%%%%%%%%%%%%

\section{ Near Light Front Old Fashioned Perturbation Theory}
%%%%%%%%%%%%%%%%%%%%%%%%%%%%%%%%%%%%%%%%%%%%%%%%%%%%%%%%%%%
\subsection{Continuum version}
%%%%%%%%%%%%%%%%%%%%%%%%%%%%%%%%%%%%%%%%%%%%
\subsubsection{$\phi^3$ theory}
%%%%%%%%%%%%%%%%%%%%%%%%%%%%%%%%%%%%%%%%%%%%
For the $\phi^3$ self-energy, we have, using the formula (\ref{nlcpt}) 
from the Appendix A
\begin{eqnarray}
\Sigma(p^2) &&= { 1 \over 2} {\lambda^2} \int_{- 
\infty}^{+ \infty} {dq_- d^2 q^\perp \over (2 \pi)^3} 
\Big (
{ 1 \over { 1 \over \eta^2} (E_{on}(p) - E_{on}(q) - E_{on}(p-q)) + 
i \epsilon} \nonumber \\ 
&& ~~~~~~-{ 1 \over { 1 \over \eta^2} 
(E_{on}(p) + E_{on}(q) + E_{on}(p-q)) -i \epsilon} \Big ) \\
&& =\Sigma_{I}(p^2) + \Sigma_{II}(p^2).
\end{eqnarray}
The two contributions  correspond to two different time orderings 
(Figs. 1a and 1b) in 
old fashioned perturbation theory. 

Let us now take the $\eta \rightarrow 0$ limit of these expressions. 
First consider $\Sigma_{I}(p^2)$. 
We have, $ \lim_{\eta \rightarrow 0}$ $ E_{on}(q) = \mid q \mid + {\eta^2
({m^2 + (q^\perp)^2)}\over 2 \mid q \mid} + ... $. 
Without loss of generality, we shall set $p_- > 0$. Then we get
\begin{eqnarray}
\Sigma_I(p^2) &&= { 1 \over 2} { \lambda^2 \over (2 \pi)^3}
\int_{- \infty}^{+\infty} dq_- d^2 q^\perp ~{ 1 \over 2 \mid q_- \mid}
~{ 1 \over 
2 \mid p_- - q_- \mid} \nonumber \\
&&~~~~{ 1 \over 
{ p_- \over \eta^2} - { \mid q_- \mid \over \eta^2} - { \mid p_- - q_- \mid
\over \eta^2}+ {m^2 + (p^\perp)^2 \over 2 p_-} - { m^2 + 
(q^\perp)^2\over 2 \mid q_- \mid} 
- { m^2 + (p^\perp - q^\perp)^2 \over 2 \mid p_- - q_- \mid}}.
\end{eqnarray}
 
Now we have to distinguish various regions. For 
$q_- > 0 $, $p_- - q_- > 0$, we get
\begin{eqnarray}
\Sigma_I(p^2) &&= { 1 \over 2} { \lambda^2 \over ( 2 \pi)^3} \int_0 ^{p_-} dq_-
\int_{- \infty}^{+ \infty} d^2q^\perp ~{ 1 \over q_-} ~{ 1 \over p_- - q_-}~ 
{ 1 \over { m^2 +(p^\perp)^2\over 2 p_-} - {m^2 +(q^\perp)^2 \over 
2 ( q_-)} -{m^2 + (p^\perp - q^\perp)^2 \over 2(p_- - q_-)}}
+ {\cal O}(\eta^2) \nonumber \\
\end{eqnarray}
which agrees with the light front answer.
For $ q_- > 0$, $ p_- - q_- < 0$, the amplitude scales as $ \eta^2 $ 
which vanishes as $ \eta \rightarrow 0 $. For $ q_- < 0$, 
$p_- - q_- > 0$  the amplitude again  scales as $\eta^2 $ 
and thus vanishes also. 

Next we consider $\Sigma_{II}(p^2)$. In the limit $ \eta \rightarrow 0 $ , we
 get
\begin{eqnarray}
\Sigma_{II}(p^2) &&= - { 1 \over 2}  \lambda^2 
\int_{- \infty}^{+ \infty} {dq_- d^2 q^\perp \over ( 2 \pi)^3} 
~{ 1 \over 2 \mid q_ - \mid }
~{ 1 \over 2 \mid p_- -q_-\mid} \nonumber \\
&&~~~{ 1 \over { 1 \over \eta^2}\Big (
p_- + \mid q_- \mid + \mid p_- - q_- \mid \Big ) + {m^2+ (p^\perp)^2 \over 2 p_-}
+ {m^2 + (q^\perp)^2 \over 2 \mid q_- \mid} + {m^2 
+ (p^\perp - q^\perp)^2 \over 2 \mid p_- - q_- \mid } }.
\end{eqnarray}
For the three cases namely, (a) $ q_- > 0$, $ p_- - q_- > 0$, (b)
$q_- > 0$, $p_- - q_- > 0$, and (c) $q_-< 0, p_- - q_-> 0$, we find that 
$ \Sigma_{II}(p^2) $ scales as $\eta^2 $ which vanishes in the limit.  

Thus we observe that for $ \phi^3$ self-energy, for finite $\eta$ there are
two time ordered diagrams. As $ \eta \rightarrow 0$ the "backward moving"
diagram vanishes as $\eta^2$ and we get the light front perturbation theory
answer. It is important to note that for any value of $\eta$, the sum of the
two contributions should be independent of $\eta$ as dictated by Lorentz
invariance. However, it is sufficient for our purposes to 
show $\eta$ independence in the limit $ \eta \rightarrow 0$.  
%%%%%%%%%%%%%%%%%%%%%%%%%%%%%%%%%%%%%%%
\subsubsection{$\phi^4$ theory}
%%%%%%%%%%%%%%%%%%%%%%%%%%%%%%%%%%%%%%%
For the scattering in $\phi^4$ theory, we have two time ordered 
diagrams (Figs. 3a and 3b). 
First consider Fig. 3a. We denote $p_- = p_{1-} - p_{3-}$.
We have 
\begin{eqnarray}
M_{fi(I)} &&= { 1 \over 2} {\lambda^2} \int_{- 
\infty}^{+ \infty} {dq_- d^2 q^\perp \over (2 \pi)^3}~ 
{ 1 \over 2 E_{on}(q)} ~{ 1 \over 2 E_{on}(p-q)} \nonumber \\
&& ~~~~
{ 1 \over { 1 \over \eta^2} (E_{on}(p_{1-}) - E_{on}(p_{3-})
 - E_{on}(q) - E_{on}(p-q)) + 
i \epsilon} .  
\end{eqnarray}
Now consider the limit  $ \eta \rightarrow 0$. 
The energy denominator becomes
\begin{eqnarray}
{ 1 \over { 1 \over \eta^2}\Big ( p_- - \mid q_-\mid  - \mid p_- - q_- \mid 
\Big )+ {m^2 + (p_1^\perp)^2 \over 2 p_1^+} - {m^2 + (p_3^\perp)^2 
\over 2 p_3^+}
- {m^2 + (q^\perp)^2 \over2  \mid q_- \mid}
- { m^2 + (p^\perp - q^\perp)^2 \over 2 \mid p_- - q_- \mid}}.
\end{eqnarray}
The analysis proceeds as in the case of $ \Sigma_{I}(p^2)$. We get a 
non-vanishing contribution which matches the light front perturbation 
theory answer. For $E_{on}(p_{1-}) < E_{on}(p_{3-})$, i.e., $p_{3-} > p_{1-}$,
the analysis proceeds as in the case of $ \Sigma_{II}(p^2)$ and the 
contribution vanishes as $ \eta^2$. 

Next consider $M_{fi(II)}$ (Fig. 3b). For   
$E_{on}(p_{1-}) > E_{on}(p_{3-})$, i.e., $p_{3-} < p_{1-}$, the analysis 
proceeds as in the case of $ \Sigma_{II}(p^2)$ and the contribution vanishes. 
For the case
$E_{on}(p_{1-}) < E_{on}(p_{3-})$, i.e., $p_{3-} > p_{1-}$, the analysis
proceeds as in the case of $ \Sigma_{I}(p^2)$ and we get a non-vanishing 
contribution that agrees with the light front answer.
%%%%%%%%%%%%%%%%%%%%%%%%%%%%%%%%%%%%%%%%%%%%%%%%%%%%%%%%
\subsection{Discretized version}
%%%%%%%%%%%%%%%%%%%%%%%%%%%%%%%%%%%%%%%%%%%%%%%%%%%%%%%%
\subsubsection{$\phi^3$ theory}
%%%%%%%%%%%%%%%%%%%%%%%%%%%%%%%%%%%%%%%%%%%%%%%%%%%%%%%%
Let us consider the first term of $ \phi^3$ self energy diagram (Fig. 1a).
Restricting the longitudinal coordinate to a finite interval, 
we obtain
\begin{eqnarray}
\Sigma_{I}(p^2) &&= { 1 \over 2} \lambda^2
{ 1 \over 2 L}\sum_n \int { d^2 q^\perp \over (2 \pi)^2}
{ 1 \over 2\sqrt{({n \pi \over L})^2 + \eta^2((q^\perp)^2 + m^2)}}
\nonumber \\
&& ~~~~~{ 1 \over 2\sqrt{({(j-n) \pi \over L})^2 + 
\eta^2((p^\perp- q^\perp)^2 + m^2)}} 
\nonumber \\
&&~~~{ 1 \over { 1 
\over \eta^2}(E_{i} - E_{I}) + i \epsilon}
\end{eqnarray}
where the energy of the initial (i) and intermediate (I) state is given by 
\begin{eqnarray}
E_i && = \sqrt{ ({j \pi \over L})^2 + \eta^2(
(p^\perp)^2 + m^2)}, \nonumber \\
E_{I}&& = \sqrt{ ({n \pi \over L})^2 + \eta^2((q^\perp)^2 + m^2)}
+ \sqrt{ ({(j-n) \pi \over L})^2 + \eta^2((p^\perp - q^\perp)^2 + m^2}) 
\end{eqnarray}
and the 
discretized longitudinal momenta are
\begin{eqnarray}
q_- = {n\pi \over L},\;\;p_- = {j\pi \over L},\;\;n,j = 0,\pm1,\pm 2,\dots
\end{eqnarray}
For $ j, n \neq 0, $ as $\eta \rightarrow 0$, we get the result independent of 
$ \eta$ and $L$.

For $ n > j$, the amplitude vanishes as $ \eta^2 L^2$ for fixed $L$. 
For $ n=j =0$, 
the amplitude diverges as $ { 1 \over \eta L}$.

For Fig. 1b, we have   
\begin{eqnarray}
\Sigma_{II}(p^2) &&= -{ 1 \over 2} \lambda^2
{ 1 \over 2 L}\sum_n \int { d^2 q^\perp \over (2 \pi)^2}
{ 1 \over 2\sqrt{({n \pi \over L})^2 + \eta^2((q^\perp)^2 + m^2)}}
\nonumber \\
&& ~~~~~{ 1 \over 2\sqrt{({(j-n) \pi \over L})^2 + 
\eta^2((p^\perp- q^\perp)^2 + m^2)}} 
\nonumber \\
&&~~~{ 1 \over { 1 
\over \eta^2}(E_{i} + E_{I}) - i \epsilon} .
\end{eqnarray}
For $ j,n \neq 0$, as $\eta \rightarrow 0$, the amplitude vanishes as $\eta^2
L^2$. For $n=j=0$, the amplitude diverges as ${ 1 \over L \eta}$. It is 
not difficult to understand the origin of this divergence. We have already seen 
that there is no dynamical scalar zero mode on the light front and thus the 
sum over intermediate states cannot include this mode. On the other hand, for 
arbitrarily small but non-zero $\eta$ (space-like quantization) there is a 
dynamical zero mode in the sum over intermediate states. By requiring this state 
to exist in the limit we are not approaching the light front theory but some 
peculiar (divergent) regime of the space-like theory. The light-front theory 
has its own mechanisms (constraints for zero modes) to replace this ``missing'' 
dynamical mode.  
%%%%%%%%%%%%%%%%%%%%%%%%%%%%%%%%%%%%%%%%%%%%%%%%%%%%%%%%%%%%%
\subsubsection{$\phi^4$ theory}
%%%%%%%%%%%%%%%%%%%%%%%%%%%%%%%%%%%%%%%%%%%%%%%%%%%%%%%%%%%%%  
The scattering amplitude in the discretized form for Fig. 3a reads  
\begin{eqnarray}
M_{fi(I)} &&= {1 \over 2}\lambda^2{1 \over {2L}} 
\sum_n \int { d^2 q^\perp \over (2 \pi)^2}
{ 1 \over 2\sqrt{({n \pi \over L})^2 + \eta^2((q^\perp)^2 + m^2)}}
\nonumber \\
&& ~~~~~{ 1 \over 2\sqrt{({(j-k-n) \pi \over L})^2 + 
\eta^2((p_1^\perp- p_3^\perp-q^\perp)^2 + m^2)}} 
\nonumber \\
&&~~~{ 1 \over { 1 
\over \eta^2}(E_{i} - E_{I}) + i \epsilon}
\end{eqnarray}
where 
\begin{eqnarray}
E_i && = \sqrt{ ({j \pi \over L})^2 + \eta^2(
(p_1^\perp)^2 + m^2)} - \sqrt{ ({k \pi \over L})^2 + \eta^2(
(p_3^\perp)^2 + m^2)}, \nonumber \\
E_{I}&& = \sqrt{ ({n \pi \over L})^2 + \eta^2((q^\perp)^2 + m^2)}
+ \sqrt{ ({(j-k-n) \pi \over L})^2 + \eta^2((p_1^\perp - p_3^\perp-q^\perp)^2 
+ m^2}) 
\end{eqnarray}
and the 
discretized longitudinal momenta are
\begin{eqnarray}
q_- = {\pi \over L}n,\;\;p_{1-} = {\pi \over L}j,\;\;p_{3-} = {\pi \over L}k,
\;\;n,j = 0,\pm1,\pm 2,\dots
\end{eqnarray}
For Fig. 3b, we get
\begin{eqnarray}
M_{fi(II)} &&= {1 \over 2}\lambda^2{1 \over {2L}} 
\sum_n \int { d^2 q^\perp \over (2 \pi)^2}
{ 1 \over 2\sqrt{({n \pi \over L})^2 + \eta^2((q^\perp)^2 + m^2)}}
\nonumber \\
&& ~~~~~{ 1 \over 2\sqrt{({(k-j-n) \pi \over L})^2 + 
\eta^2((p_3^\perp- p_1^\perp-q^\perp)^2 + m^2)}} 
\nonumber \\
&&~~~{ 1 \over { 1 
\over \eta^2}(E_{i} - E_{I}) + i \epsilon},
\end{eqnarray}
where 
\begin{eqnarray}
E_i && = \sqrt{ ({k \pi \over L})^2 + \eta^2(
(p_3^\perp)^2 + m^2)} - \sqrt{ ({j \pi \over L})^2 + \eta^2(
(p_1^\perp)^2 + m^2)}, \nonumber \\
E_{I}&& = \sqrt{ ({n \pi \over L})^2 + \eta^2((q^\perp)^2 + m^2)}
+ \sqrt{ ({(k-j-n) \pi \over L})^2 + \eta^2((p_3^\perp - p_1^\perp-q^\perp)^2 
+ m^2}).
\end{eqnarray}
As in the case of $\phi^3$ theory, for $j-k \neq 0$, $ n \neq 0$ there is no 
problem, but for
$j-k=0$, $n=0$, we run into divergences.
%%%%%%%%%%%%%%%%%%%%%%%%%%%%%%%%%%%%%%%%%%%%%%%%%%%%%%%%%%%%%%
\section{Infinite Momentum Frame Approach}
%%%%%%%%%%%%%%%%%%%%%%%%%%%%%%%%%%%%%%%%%%%%%%%%%%%%%%%
\subsection{Continuum version}
%%%%%%%%%%%%%%%%%%%%%%%%%%%%%%%%%%%
\subsubsection{$\phi^3$ theory}
%%%%%%%%%%%%%%%%%%%%%%%%%%%%%%%%%%%%%%%%%%%%
For the $\phi^3$ self energy, (Figs. 1a and 1b), using the rules of old 
fashioned perturbation theory, we obtain
\begin{eqnarray}
\Sigma(p^2) &&= { 1 \over 2} {\lambda^2} \int_{- 
\infty}^{+ \infty} {d^3q  \over (2 \pi)^3} 
{ 1 \over 2 E_{q}} { 1 \over 2 E_{p-q}}
\Big( 
{ 1 \over  E_{p} - E_{q} - E_{p-q} + 
i \epsilon} \nonumber \\ 
&& ~~~~~~-{ 1 \over  E_p + E_{q} + E_{p-q} -i \epsilon} \Big ).  \\
&& =\Sigma_{I}(p^2) + \Sigma_{II}(p^2).
\end{eqnarray}
Here $ E_p = \sqrt{p^2 +(p^\perp)^2 + m^2}$. For ease of notation we have
denoted the third component of the three-vector ${\bf p}$ as $p$. 
The two contributions  correspond to two different time orderings in 
old fashioned perturbation theory. 
To facilitate the infinite momentum limit, we parametrize the internal 
momenta as follows: $ {\bf q} = (x p , q^\perp)$, $ {\bf p-q} = ((1 -x) p ,
 p^\perp - q^\perp) $. It is important to note that the range of $x$ is 
$ - \infty < x < + \infty$.  
Now $ {d^3q \over 2 E_q}= {{p dx d^2 q^\perp}\over 2 E_q}$. 
Let us now take the infinite momentum, $p \rightarrow \infty$ 
limit of these expressions. 
It follows that ${ p dx \over 2 E_q} \rightarrow
 { 1 \over 2 }{ dx \over \mid x \mid}$, 
$E_q \rightarrow \mid x \mid p + { m^2 + (q^\perp)^2 \over 2 \mid  x \mid p}
$.
 
First consider $\Sigma_{I}(p^2)$. 
We get
\begin{eqnarray}
\Sigma_I(p^2) &&= { 1 \over 2} { \lambda^2 \over (2 \pi)^3}
\int_{- \infty}^{+\infty} dx d^2 q^\perp ~{ 1 \over 2 x}~{ 1 \over
2 \mid 1-x \mid p} \nonumber \\
&&~~~~{ 1 \over 
 p(1-   \mid x \mid   -  \mid
1-x \mid )+ {m^2 + (p^\perp)^2 \over 2 p} - { m^2 + 
(q^\perp)^2\over 2 \mid x \mid p} 
- { m^2 + (p^\perp - q^\perp)^2 \over 2 \mid 1- x \mid p}}.
\end{eqnarray}
 
Now we have to distinguish various regions. For 
$x \geq 0 $, $1-x \geq 0$, we get
\begin{eqnarray}
\Sigma_I(p^2) = { 1 \over 2} { \lambda^2 \over ( 2 \pi)^3} \int_0 ^{1} dx
{ 1 \over { m^2 +(p^\perp)^2\over 2} - {m^2 + (q^\perp)^2 \over 2 x} 
-{m^2 + (p^\perp - q^\perp)^2\over 2(1-x)}}
\end{eqnarray}
which agrees with the light front answer.
For $ x > 0$, $ 1-x < 0$, the amplitude scales as $ { 1 \over p^2} $ 
which vanishes as $ p \rightarrow \infty$. For $ x < 0$, 
$1-x > 0$  the amplitude again  scales as ${ 1 \over p^2} $  
{\tt and} vanishes in the limit. 

Next we consider $\Sigma_{II}(p^2)$. In the limit $ p \rightarrow \infty $ , 
we get
\begin{eqnarray}
\Sigma_{II}(p^2) &&= - { 1 \over 2} { \lambda^2 \over  ( 2 \pi)^3} 
\int_{- \infty}^{+ \infty} dx ~{ 1 \over 2 \mid x \mid }
~{ 1 \over 2 \mid 1-x \mid p} \nonumber \\
&&~~~{ 1 \over \Big (
p(1 + \mid x \mid  + \mid 1-x  \mid )+ {m^2+ (p^\perp)^2 \over 2 p}
+ {m^2 + (q^\perp)^2 \over 2 \mid x \mid p} + {m^2 
+ (p^\perp - q^\perp)^2 \over 2 \mid 1-x \mid p} \Big )}.
\end{eqnarray}
For the three cases namely, (a) $ x > 0$, $ 1-x > 0$, (b)
$x > 0$, $1-x < 0$, and (c) $x< 0, 1-x> 0$, we find that 
$ \Sigma_{II}(p^2) $ scale as ${ 1 \over p^2} $ which vanishes in the limit.

Thus in old fashioned perturbation theory in the infinite momentum 
limit ($ p \rightarrow \infty$),
the $``$backward going diagram" vanishes as $ { 1 \over p^2}$ in accordance 
with  Weinberg's results\cite{wein}.   
%%%%%%%%%%%%%%%%%%%%%%%%%%%%%%%%%%%%%%%
\subsubsection{$\phi^4$ theory}
%%%%%%%%%%%%%%%%%%%%%%%%%%%%%%%%%%%%%%%
For the scattering in $\phi^4$ theory, we have two time ordered diagrams
(Figs. 3a and 3b). 
Let us start with the first one.  We denote 
$p= p_{1} - p_{3}$.
We have 
\begin{eqnarray}
 M_{fi(I)} &&= { 1 \over 2} {\lambda^2} \int_{- 
\infty}^{+ \infty} {dq d^2 q^\perp \over (2 \pi)^3} 
~{ 1 \over 2 E_{q}} ~{ 1 \over 2 E_{p-q}} \nonumber \\
&& ~~~~
{ 1 \over  (E_{p_{1}} - E_{p_{3}}
 - E_{q} - E_{p-q}) + 
i \epsilon}  .
\end{eqnarray}
Now consider the limit $ p \rightarrow  \infty$.  
The energy denominator becomes
\begin{eqnarray}
&&{ 1 \over p \big ( (x_1 - x_3) - \mid x \mid  - \mid x_1 - x_3 -x \mid \big ) 
+ {m^2 + (p_1^\perp)^2 \over 2 x_1p} 
- {m^2 + (p_3^\perp)^2 \over 2 x_3 p}
- {m^2 + (q^\perp)^2 \over 2 \mid x\mid p}
- { m^2 + (p^\perp - q^\perp)^2 \over 2\mid x_1 - x_3 - x \mid p}}. 
\nonumber \\
&&
\end{eqnarray}
The analysis proceeds as in the case of $ \Sigma_{I}(p^2)$. We get a 
non-vanishing contribution which matches 
the corresponding part of the total 
light front answer, Eq. (\ref{lff1}), for
$E_{p_{1}} > E_{p_{3}}$, i.e., $x_1 > x_3$.
 For $E_{p_{1}} < E_{p_{3}}$, i.e., $x_1 < x_3$,
the analysis proceeds as in the case of $ \Sigma_{II}(p^2)$ and the 
contribution vanishes as $ { 1 \over p^2}$. 

Next consider $M_{fi(II)}$.
\begin{eqnarray}
M_{fi(II)} = -{ 1 \over 2} {\lambda^2} \int_{- 
\infty}^{+ \infty} {dq d^2 q^\perp \over (2 \pi)^3} 
{ 1 \over  (E_{p_{1}} - E_{p_{3}}
 + E_{q} + E_{p-q}) - 
i \epsilon}  .
\end{eqnarray}

For   
$E_{p_{1}} > E_{p_{3}}$, i.e., $x_{1} > x_{3}$, the analysis 
proceeds as in the case of $ \Sigma_{II}(p^2)$ and the contribution vanishes. 
For the case
$E_{p_{1}} < E_{p_{3}}$, i.e., $x_{3} > x_{1}$, the analysis
proceeds as in the case of $ \Sigma_{I}(p^2)$ and we get a non-vanishing 
contribution that agrees with the corresponding contribution of the 
light front answer, Eq. (\ref{lff2}). Thus the total result, Eq. (\ref{lft}), is
obtained.
%%%%%%%%%%%%%%%%%%%%%%%%%%%%%%%%%%%%%%%%%%%%%%%%%%%%%%%%
\subsection{Discretized version}
%%%%%%%%%%%%%%%%%%%%%%%%%%%%%%%%%%%%%%%%%%%%%%%%%%%%%%%
\subsubsection{$\phi^3$ theory}
%%%%%%%%%%%%%%%%%%%%%%%%%%%%%%%%%%%%%%%%%%%%%%%%%%%%%%%
As before, we restrict the longitudinal coordinate to a 
finite interval. Specifically, we set $ -L < x^3 < +L$. The longitudinal 
momenta
$ q^3 \rightarrow q^3_n = {  \pi \over  L} n$, $ n=0, \pm1, \pm 2, \ldots $. 
The field operator becomes
\begin{eqnarray}
\phi(x) = { 1 \over \sqrt{2L}} \sum_n \int d^2 q^\perp 
{ 1 \over \sqrt{2 \omega_n}} \Big [ 
a_n(q^\perp) e^{ -i \omega_n t + i { n \pi x^3  \over L} + i q^\perp \cdot 
x^\perp } + a^\dagger_n(q^\perp) 
e^{ i \omega_n t - i { n \pi x^3  \over L} - i q^\perp \cdot x^\perp} \Big ].
\end{eqnarray}
Let us consider the $\phi^3$ self energy. 
For the external momentum we set $p=({j \pi \over L}, p^\perp)$.  We get
\begin{eqnarray}
\Sigma(p^2) &&= { 1 \over 2} { \lambda^2 } { 1 \over 2L} \sum_n
\int{ d^2 q^\perp \over (2 \pi)^2} 
{ 1 \over 2 \sqrt{({n \pi \over L})^2 + (q^\perp)^2 + m^2}} 
{ 1 \over 2 \sqrt{({(j-n) \pi \over L})^2 + (p^\perp- q^\perp)^2 + m^2}} 
\nonumber \\
&& \Big (
{ 1 \over \sqrt{ ({j \pi \over L})^2 + (p^\perp)^2 + m^2}
-  \sqrt{ ({n \pi \over L})^2 + (q^\perp)^2 + m^2}
- \sqrt{ ({(j-n) \pi \over L})^2 + (p^\perp - q^\perp)^2 + m^2}}
\nonumber \\
&&~~ -
{ 1 \over \sqrt{ ({j \pi \over L})^2 + (p^\perp)^2 + m^2}
+  \sqrt{ ({n \pi \over L})^2 + (q^\perp)^2 + m^2}
+  \sqrt{ ({(j-n) \pi \over L})^2 + (p^\perp - q^\perp)^2 + m^2}}\Big).
\nonumber \\
&&
\end{eqnarray}
If we take the continuum limit,  then 
$ { 1 \over 2L} \sum_n \rightarrow { dq \over 2 \pi}$
and we obtain the result of the previous subsection. 
Then, taking the infinite momentum limit, the second contribution 
drops out and 
we get the light front answer from the first contribution alone. 

Suppose one takes the limit $ L \rightarrow 0$, which is 
the opposite of the continuum limit $L \rightarrow \infty$. This is an attempt to simulate
DLCQ results in a space like box. We do not expect the result to agree with 
the continuum limit of DLCQ which agrees with covariant perturbation 
theory results.

For $ n \neq 0, j \neq 0, n < j$, in the limit $ L \rightarrow 0$, 
the amplitude becomes independent of $L$. For $ j=n=0$, the amplitude 
diverges like $ { 1 \over L}$. For $ n > j$ the amplitude vanishes like 
$ L^2$. But none of these results have anything to do with either 
continuum or DLCQ results. 
%%%%%%%%%%%%%%%%%%%%%%%%%%%%%%%%%%%%%%%%%%%%%%%%%%%%%%%%%%
\subsubsection{$\phi^4$ theory}
%%%%%%%%%%%%%%%%%%%%%%%%%%%%%%%%%%%%%%%%%%%%%%%%%%%%%%%%%%
The scattering amplitude for Fig. 3a is given by
\begin{eqnarray}
M_{fi(I)} &&= { 1 \over 2} \lambda^2 { 1 \over 2L} \sum_n
\int{ d^2 q^\perp \over (2 \pi)^2} 
{ 1 \over 2\sqrt{({n \pi \over L})^2 + (q^\perp)^2 + m^2}} \nonumber \\
&&{ 1 \over 2\sqrt{({(j-k-n) \pi \over L})^2 + (p_1^\perp-p_3^\perp- q^\perp)^2 
+ m^2}}{ 1 \over {E_i - E_I +i\epsilon}},
\end{eqnarray}
where
\begin{eqnarray}
E_i &=& \sqrt{ ({j \pi \over L})^2 + (p_1^\perp)^2 + m^2}
- \sqrt{ ({k \pi \over L})^2 + (p_3^\perp)^2 + m^2} \nonumber \\
E_I &=&   \sqrt{ ({n \pi \over L})^2 + (q^\perp)^2 + m^2}
+ \sqrt{ ({(j-k-n) \pi \over L})^2 + (p_1^\perp - p_3^\perp -q^\perp)^2 + m^2}.
\end{eqnarray}
and the discretized longitudinal momenta are 
\begin{eqnarray}
q_n^3 = {\pi \over L}n,\;\;p_{1j}^3 = {\pi \over L}j,\;\;p_{3k} ^3 = 
{\pi \over L}k,\;\;j,k,n = \pm 1,\pm2,\dots.
\end{eqnarray}
For Fig. 3b we find 
\begin{eqnarray}
M_{fi(II)} &&=  { 1 \over 2} \lambda^2 { 1 \over 2L} \sum_n
\int{ d^2 q^\perp \over (2 \pi)^2} 
{ 1 \over 2\sqrt{({n \pi \over L})^2 + (q^\perp)^2 + m^2}} \nonumber \\ 
&& { 1 \over 2\sqrt{({(k-j-n) \pi \over L})^2 + (p_3^\perp-p_1^\perp- 
q^\perp)^2 + m^2}} 
{1 \over {E_i - E_I +i\epsilon}},
\end{eqnarray} 
where
\begin{eqnarray}
E_i &=&  \sqrt{ ({k \pi \over L})^2 + (p_3^\perp)^2 + m^2}
- \sqrt{ ({j \pi \over L})^2 + (p_1^\perp)^2 + m^2} \nonumber \\
E_I &=&  \sqrt{ ({n \pi \over L})^2 + (q^\perp)^2 + m^2}
+  \sqrt{ ({(k-j-n) \pi \over L})^2 + (p_3^\perp - p_1^\perp -q^\perp)^2 + m^2}.
\end{eqnarray}

As in the case of $\phi^3$ theory, for $j-k \neq 0$, $n\neq 0$, there is no  
problem but for $j-k=0$, and $n=0$, we run into divergences.  
%%%%%%%%%%%%%%%%%%%%%%%%%%%%%%%%%%%%%%%%%%%%%%%%%%%%%%%%%%%
\section{Summary and Conclusions}
%%%%%%%%%%%%%%%%%%%%%%%%%%%%%%%%%%%%%%%%%%%%%%%%%%%%%%%%%%%
In this work, we have studied continuum and discretized versions of the 
time ordered (``old fashioned '') perturbation theory in the light-front, 
near light front and infinite 
momentum frameworks, applied to scalar field theory self-energy and 
scattering amplitudes. We have recalled important features of 
the covariant perturbation theory, namely that when Feynman amplitudes are 
rewritten in terms of the light front variables and the contour integration 
in the light-front energy complex plane is performed, the Feynman amplitudes 
reduce back to the continuum light front answers \cite{ChMa}. Also, as stressed 
already by Weinberg in 1966 \cite{wein}, the light front perturbation theory 
(old-fashioned perturbation theory in a reference frame with  
"infinite-momentum" 
at that time) is more economical in the sense that  
one does not need to introduce Feynman 
parameters to combine propagators in the corresponding integrals, and the 
four-dimensional Euclidean integration is replaced by a two-dimensional one. 
Feynman parameters appear in the light front formulation naturally as 
light front longitudinal momentum fractions.

More specifically, after demonstrating that the continuum light front 
perturbation theory has no problem with zero modes and its results agree with 
the covariant results, we have analyzed the 
continuum limit of the light-front 
perturbation theory formulated in a finite volume with periodic fields (DLCQ 
method). This investigation was motivated by  claims  
\cite{pol} that DLCQ is ill-defined since it is divergent when 
formulated as a limit of the space-like quantization on a hypersurface close 
to the light front\cite{mtp}. In this connection, we have first shown that the DLCQ 
perturbation theory is consistent, because parts of the perturbative amplitudes 
due to the effective interactions induced by the constrained zero mode vanish 
in the infinite-volume limit and the covariant results are reproduced. Second,  
when one considers the light front limit ($\eta \rightarrow 0$) of the near light 
front discretized amplitudes, the zero-mode contribution indeed diverges for 
fixed box length. But this disagrees with the light front answer and actually 
cannot tell anything about the light front zero modes. The point is that 
the light front zero mode is not dynamical in the scalar theory and thus 
it is not present in the 
complete set of intermediate states. By letting $\eta \rightarrow 0$ one 
is forcing the dynamical space-like zero mode to exist on the light front 
which leads to an incorrect, diverging amplitude. Recall
the light front 
dispersion relation for a free particle (dynamical quantum) $p^- = (m^2 + (p^
\perp)^2)(p^+)^{-1}$ which gives divergent energy for a mode with $p^+ = 0$ 
(and non-zero numerator). 
It is remarkable how the light front theory copes with this problem: most  
of the light front zero modes are non-dynamical, i.e. constrained variables.  
Thus the above dispersion relation is not applicable. A few known 
dynamical zero modes are massless and have vanishing $p^\perp$ (global zero 
modes). The dispersion relation is again not applicable and one has to 
treat these modes in a different manner (in terms of zero-mode coherent states, 
e.g. \cite{lm}.) 

In other words, the $\eta \rightarrow 0$ limit does not lead us to the light 
front theory but to a peculiar non-covariant regime of the  
discretized space-like theory. 

On the other hand, the continuum version of the near light front old fashioned 
perturbation theory reproduces the light front answers (which agree with the 
covariant ones). But since this formulation has no particular advantages there 
is no real reason to use it in practical calculations.

In the continuum version of old fashioned perturbation theory in the 
infinite momentum frame, there is  no problem
in perturbation theory. In the box, the zero mode contribution scales as ${ 1
\over L}$ whereas nonzero modes scales independent of $L$. Thus in
perturbation theory zero modes decouple in the continuum 
limit ($L \rightarrow \infty$).

However, if one reinterprets the boost as the equal time box length $L_{et}
\rightarrow 0$ as is done in M-theory literature (to simulate DLCQ for
finite $L$), zero mode contributions
diverge. In this limit, however, there are  
infinite number of terms in the DLCQ Hamiltonian.

We would like to emphasize that our conclusions are based on the 
careful analysis of the {\it scalar} field theories. This choice was motivated 
by the discussion in the M-theory literature \cite{pol}, where one-loop 
scattering in ${\lambda \over {4!}}\phi^4$ scalar theory was analyzed in 
the near light front framework. It would be very instructive to perform an 
analysis, similar to that done in the present work, within a theory with 
fermions, and in particular in a gauge theory. This would help to clarify  
some unresolved issues   
in the light front literature
\cite{mb}.  For example, some authors, 
using DLCQ perturbation theory, found unpleasant quadratic divergences 
within the 3+1 dimensional light-front Yukawa model. However, the continuum 
limit was not studied in their work.

In conclusion, our answer to the question, raised in the paper  
\cite{pol} : Does it makes sense to put a quantum system to a 
light-like box ?  
is: {\it yes, it does}. Light-like compactification is feasible and DLCQ   
is consistent (there is  
no problem neither with causality \cite{dipetal}). The
discretized  (compactified) formulation of the theory on the light-like 
surface does exist 
as a straightforward light front field theory, but {\it not} as a limit of 
a space-like compactification.

%%%%%%%%%%%%%%%%%%%%%%%%%%%%%%%%
\acknowledgements
This work was supported in part by the U.S. Department 
of Energy, Grant No. DE-FG02-87ER40371, Division of High Energy and 
Nuclear physics, by the NSF Grant No. INT-9515511, by the VEGA Grant No. 
2/5085/1998 and by the International Institute of 
Theoretical and Applied Physics, Iowa State University, Ames, Iowa, U.S.A.  
%%%%%%%%%%%%%%%%%%%%%%%%%%%%%%%%%%%%%%%%%%%%%%%%%%%%%%%%%%%%
\appendix
%%%%%%%%%%%%%%%%%%%%%%%%%%%%%%%%%%%%%%%%%%%%%%%%%%%%%%%%%%%
\section{Field Theory in Near Light Front Coordinates}
%%%%%%%%%%%%%%%%%%%%%%%%%%%%%%%%%%%%%%%%%%%%%%%%%%%%%%%%%%%%
\subsection{Kinematics}
%%%%%%%%%%%%%%%%%%%%%%%%%%%%%
Consider the set of coordinates
\begin{eqnarray}
x^+ &&= { 1 \over \sqrt{2}} (x^0 + x^3 ) + { 1 \over 2 } \eta^2 {1 \over
\sqrt{2}} (x^0 - x^3) \nonumber \\
x^- &&= { 1 \over \sqrt{2}} (x^0 -x^3) \nonumber \\
x^\perp &&=(x^1,x^2).
\end{eqnarray}
We shall take $x^+$ to be the time variable. Then $x^-$ is a longitudinal
coordinate.
The metric tensor in the new coordinate system is given by
\begin{eqnarray}
{\tilde g}_{\mu \nu} = \left[ \begin{array}{cccc} 0 & 1 & 0& 0 \\
                              1 & - \eta^2 & 0 & 0 \\
                              0 & 0 & -1 & 0 \\
                              0& 0& 0& -1 \end{array} \right ], \nonumber \\
{\tilde g}^{\mu \nu} = 
\left[ \begin{array}{cccc} \eta^2 & 1 & 0& 0 \\
                              1 & 0 & 0 & 0 \\
                              0 & 0 & -1 & 0 \\
                              0& 0& 0& -1 \end{array} \right ] .
\end{eqnarray}
Thus we have,
\begin{eqnarray}
x^2 = {\tilde g}_{\mu \nu } x^\mu x^\nu = 2x^+x^- - \eta^2 (x^-)^2 -
(x^\perp)^2= {\tilde
g}^{\mu \nu} x_\mu x_\nu = \eta^2 (x_+)^2 + 2x_+x_- - (x_\perp)^2.
\end{eqnarray}
Furthermore,
\begin{eqnarray}
x_+=x^-, x_-= x^+ - \eta^2 x^-.
\end{eqnarray}
The scalar product {\tt is} $k \cdot x 
= k^+x^- + k^- x^+ - \eta^2 k^- x^-  - k^\perp \cdot
x^\perp= k_+x^+ +
k_-x^- - k_\perp \cdot x_\perp$. 
Thus $k_+$  which is conjugate to $x^+$ is the energy and $k_-$
which is conjugate to $x^-$ is the longitudinal momentum. It is important to
keep in mind that $ - \infty < k_- < + \infty$.

For an on mass-shell particle of mass $m$, $k^2 =m^2$ yields
\begin{eqnarray}
\eta^2 (k_+)^2 + 2 k_+k_- - (k^\perp)^2- m^2 =0
\end{eqnarray}
which leads to the dispersion relation
\begin{eqnarray}
 k_+ = { - k_- \pm \sqrt{(k_-)^2 + (m^2 + (k^\perp)^2)
\eta^2} \over \eta^2}.
\end{eqnarray}
For an on mass-shell particle, since $k^0 > k^3$, $ k^0 > 0 $ implies $
k_+ > 0$ and hence  
the Lorentz invariant phase space factor
\begin{eqnarray}
{d^4k \over (2 \pi)^4} 2 \pi \delta(k^2 - m^2) \theta(k_+) && 
= {dk_+ dk_- d^2 k^\perp \over (2 \pi)^3} \delta(\eta^2
(k_+)^2  + 2 k_+ k_- - (k^\perp)^2 - m^2) \theta(k_+) \nonumber \\  
&& = {dk_- d^2 k^\perp \over ( 2 \pi)^3 2 E_{on}},  
\end{eqnarray}
where $  E_{on}(k) = \sqrt{(k_-)^2 + \eta^2 ((k^\perp)^2 + m^2)}$.
%%%%%%%%%%%%%%%%%%%%%%%%%%%%%%%%%%%%%%
\subsection{Free scalar field theory}
%%%%%%%%%%%%%%%%%%%%%%%%%%%%%%%%%%%%%%
Consider the Lagrangian density
\begin{eqnarray}
{\cal L} = { 1 \over 2} \partial_\mu \phi \partial^\mu \phi - { 1 \over 2}
m^2 \phi^2  
 = { 1 \over 2} \eta^2 \partial_+ \phi \partial_+ \phi + \partial_+ \phi
\partial_- \phi - { 1 \over 2} \partial^\perp \phi \cdot \partial^\perp \phi 
- { 1 \over 2} m^2 \phi^2.
\end{eqnarray}
The equation of motion 
\begin{eqnarray}
(\partial_\mu \partial^\mu + m^2) \phi =0 
\end{eqnarray}
becomes
\begin{eqnarray}
(\eta^2 \partial_+ \partial_+ + 2 \partial_+ \partial_- - (\partial^\perp)^2
+ m^2 ) \phi=0.
\end{eqnarray}
The general solution is 
\begin{eqnarray}
\phi(x) &=& \int {d^4k \over (2 \pi)^4} f(k) 2 \pi 
\delta(k^2 - m^2) e^{i k\cdot x}, \nonumber \\
\phi(x) & = & \int {dk_- d^2 k^\perp \over (2 \pi)^3 2 E_{on}(k) } 
[ a(k) e^{i (
k_{+on}x^+ + k_- x^- - k^\perp \cdot x^\perp) } + a^*(k) e^{ -i (
  k_{+on}x^+ +k_- x^- - k^\perp \cdot x^\perp) } ].
\end{eqnarray}
In the quantum theory we have
\begin{eqnarray}
\phi(x) = 
{ 1 \over (2 \pi)^3} \int_{- \infty}^{+ \infty} { dk_- d^2 k^\perp
\over 2 E_{on}(k)} 
[ a(k) e^{i (
k_{+on}x^+ +k_- x^- - k^\perp \cdot x^\perp) } + a^\dagger(k) e^{ -i (
  k_{+on}x^++ k_- x^- - k^\perp \cdot x^\perp) } ].
\end{eqnarray}
The conjugate momentum {\tt is}
\begin{eqnarray}
\pi(x) & = & { \partial {\cal L } \over \partial \partial_+ \phi} 
=\eta^2 \partial_+ \phi + \partial_- \phi,
\end{eqnarray}
\begin{eqnarray}
\pi(x) && = - i 
{ 1 \over (2 \pi)^3} \int_{- \infty}^{+ \infty} { dk_- d^2 k^\perp
\over 2 E_{on}(k)} E_{on}(k)
[ a(k) e^{i (
k_{+on}x^+ +k_- x^- - k^\perp \cdot x^\perp) } \nonumber \\
&&~~- a^\dagger(k) e^{ -i (
  k_{+on}x^++ k_- x^- - k^\perp \cdot x^\perp) } ].
\end{eqnarray}
We have,
\begin{eqnarray}
\left [ \phi(x), \pi(y) \right ]_{x^+=y^+} = i \delta(x^- - y^-) \delta^2(x^\perp - y^\perp),
\end{eqnarray}
provided
\begin{eqnarray}
\left [ a(k), a^\dagger (k') \right ] &&= 
(2 \pi)^3 2 E_{on}(k) \delta(k_- - k'_-) \delta^2 (k^\perp - k'^\perp),
\nonumber \\ 
\left [ a(k), a(k') \right ]&&=0,
\left [ a^\dagger(k) , a^\dagger(k') \right] =0.
\end{eqnarray}
%%%%%%%%%%%%%%%%%%%%%%%%%%%%%%%%%%%%%%%%%%%%%%%%%%%%%
%\subsubsection{Hamiltonian}
%%%%%%%%%%%%%%%%%%%%%%%%%%%%%%%%%%%%%%%%%%%%%%%%%%%%%
The Hamiltonian density {\tt is}
\begin{eqnarray}
{\cal H} &&= \pi \partial_+ \phi - {\cal L} 
 = { 1 \over 2} {(\pi - \partial_- \phi)^2 \over \eta^2} + { 1 \over 2} 
\partial^\perp \phi \cdot \partial^\perp \phi + { 1 \over 2} m^2 \phi^2
\end{eqnarray}
and the Hamiltonian in the Fock representation takes the form 
\begin{eqnarray}
H = \int dx^- d^2 x^\perp {\cal H}
 = \int {dk_- d^2 k^\perp \over ( 2 \pi)^3 2 E_{on}(k)} 
{ - k_- + \sqrt{(k_-)^2 + (m^2 + (k^\perp)^2)\eta^2} \over \eta^2}
a^\dagger(k) a(k).
\end{eqnarray}
%%%%%%%%%%%%%%%%%%%%%%%%%%%%%%%%%%%%%%%%%%%%%%%%%%%%%
%\subsubsection{propagator} 
%%%%%%%%%%%%%%%%%%%%%%%%%%%%%%%%%%%%%%%%%%%%%%%%%%%%%%
The propagator is given by
\begin{eqnarray}
i S_B(x) &&= \langle 0 \mid T(\phi(x) \phi(0)) \mid 0 \rangle 
 = \theta(x^+) \langle 0 \mid \phi(x) \phi(0) \mid 0 \rangle + \theta(-x^+) 
\langle 0 \mid \phi(0) \phi(x) \mid 0 \rangle, \nonumber \\
 &&= { 1 \over (2 \pi)^3} \int {dk_- d^2 k^\perp \over 2 E_{on}(k) }
\left [ \theta(x^+) e^{-i (k_+x^+ + k_-x^- - k^\perp \cdot x^\perp )}
+ \theta(-x^+) e^{i (k_+x^+ + k_-x^- - k^\perp \cdot x^\perp )} \right ].
\nonumber \\
\end{eqnarray}
Using 
\begin{eqnarray}
\theta(x) = { 1 \over 2 \pi i} \int_{- \infty}^{+ \infty} dy e^{i yx} 
{ 1 \over y - i \epsilon}
\end{eqnarray}
and changing integration variables, we get
\begin{eqnarray}
i S_B(x) &=& { 1 \over (2 \pi)^4} \int dk_+ dk_- d^2 k^\perp 
e^{i (k_+x^++k_-x^- - k^\perp \cdot x^\perp)}{ i \over 
\eta^2 (k_+)^2 + 2 k_+ k_- - (k^\perp)^2 -m^2 + i \epsilon},
\nonumber \\ 
i S_B(x)& = & { 1 \over (2 \pi)^4} \int d^4k 
e^{i k.x}{ i \over 
k^2 -m^2 + i \epsilon}.
\end{eqnarray}
%%%%%%%%%%%%%%%%%%%%%%%%%%%%%%%%%%%%%%%%%%%%%%%%%%%%%%%%%%%%%%%%%%%
\subsection{Old fashioned perturbation theory}
%%%%%%%%%%%%%%%%%%%%%%%%%%%%%%%%%%%%%%%%%%%%%%%%%%%%%%%%%%%%%%%%%%%%

We have the perturbative formula for the S matrix:
\begin{eqnarray}
S_{fi} = \delta_{fi} - 2 \pi i \delta(p_{+(on)f} - p_{+(on)i}) T_{fi},
\end{eqnarray}
\begin{eqnarray}
T_{fi} =  \langle f \mid V_S \mid i \rangle 
+ \sum_n {\langle f \mid V_S \mid n \rangle \langle n 
\mid  V_S \mid i \rangle \over 
p_{+(on)i} - p_{+(on)n} + i \epsilon} + \ldots 
\end{eqnarray}
where the on-shell energy 
$ p_{+(on)} = { - p_- + \sqrt{(p_-)^2 + \eta^2 (m^2 + (p^\perp)^2 
)} \over \eta^2}$.
Since the longitudinal momentum $p_-
$ is conserved at the vertex, we get,
\begin{eqnarray} 
T_{fi} =  \langle f \mid V_S \mid i \rangle 
+ \sum_n {\langle f \mid V_S \mid n \rangle \langle n 
\mid  V_S \mid i \rangle \over 
{ 1 \over \eta^2}(E_{(on)
i} - E_{(on)n}) + i \epsilon} + \ldots 
\label{nlcpt}
\end{eqnarray}
where $ E_{on}(p) = \sqrt{ (p_-)^2 + \eta^2 (m^2 + (p^\perp)^2)}$. 
The sum over intermediate states $ \sum_n 
\rightarrow \int { dk_- d^2 k^\perp \over (2 \pi)^3 2 E_{on}}$.

\eject
\centerline{\bf Figures}
\vskip 1in
\begin{center}
\begin{picture}(600,200)(0,0)
\SetWidth{1.25}
\CArc(75,75)(50,30,150)
\CArc(75,125)(50,210,330)
\Line(0,100)(33,100)
\Line(117,100)(150,100)
\Text(5,107)[l]{p}
\Text(145,107)[l]{p}
\Text(80,135)[l]{q}
\Text(80,66)[l]{p-q}
\Text(80,25)[l]{(a)}
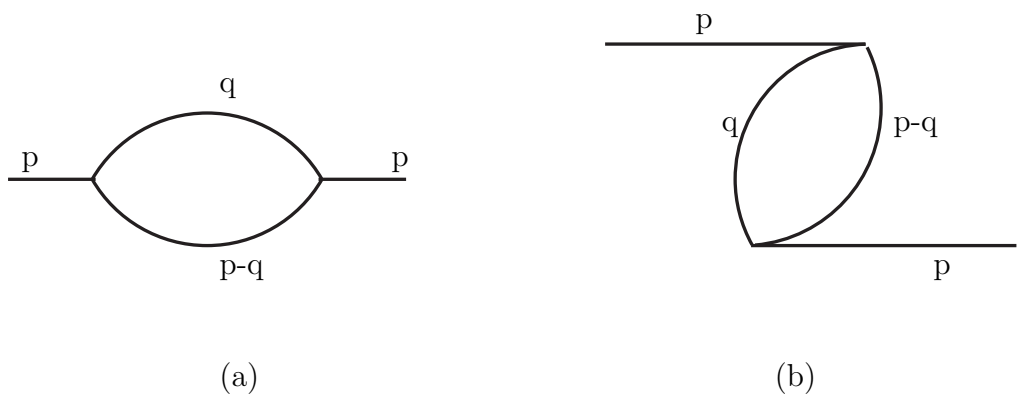
\Text(,-25)[l]{Fig. 1. $\phi^3$ self-energy diagrams in old fashioned
perturbation theory}
\CArc(325,100)(51,92,209)
\CArc(277,127)(52,275,26)
\Line(225,151)(323,151)
\Line(280,75)(380,75)
\Text(260,159)[l]{p}
\Text(350,66)[l]{p}
\Text(270,120)[l]{q}
\Text(335,120)[l]{p-q}
\Text(290,25)[l]{(b)}
\end{picture}
\begin{picture}(300,200)(0,0)
\SetWidth{1.25}
\BCirc(150,50){50}
\Line(25,50)(100,50)
\Line(200,50)(275,50)
\Text(50,60)[l]{p}
\Text(225,60)[l]{p}
\Text(150,110)[l]{q}
\Text(150,-15)[l]{p-q}
\Text(,-50)[l]{Fig. 2. $\phi^3$ self-energy diagram}
\end{picture}
\end{center} 
\eject
\vskip .25in
\begin{center}
\begin{picture}(600,300)(0,0)
\SetWidth{1.25}
\CArc(325,100)(50,90,209)
\CArc(277,127)(52,275,26)
\Line(225,150)(380,150)
\Line(225,75)(380,75)
\Text(230,158)[l]{p$_{1}$}
\Text(375,158)[l]{p$_{3}$}
\Text(230,80)[l]{p$_{2}$}
\Text(375,80)[l]{p$_{4}$}
\Text(268,115)[l]{q}
\Text(331,115)[l]{p$_{1}$-p$_{3}$-q}
\Text(290,30)[l]{(b)}
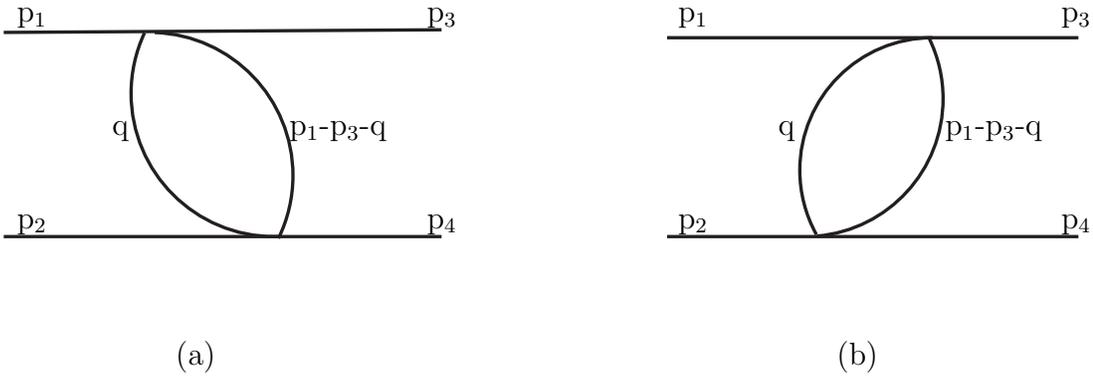
\Text(0,-5)[l]{Fig. 3. $\phi^4$ scattering diagrams in old fashioned
perturbation theory}
\CArc(30,98)(54,334,88)
\CArc(77,129)(54,155,272)
\Line(-25,152)(140,153)
\Line(-25,75)(140,75)
\Text(-20,158)[l]{p$_{1}$}
\Text(135,158)[l]{p$_{3}$}
\Text(-20,80)[l]{p$_{2}$}
\Text(135,80)[l]{p$_{4}$}
\Text(16,115)[l]{q}
\Text(83,115)[l]{p$_{1}$-p$_{3}$-q}
\Text(40,30)[l]{(a)}
\end{picture}
\end{center} 
\begin{center}
\begin{picture}(300,200)(0,0)
\SetWidth{1.25}
\CArc(150,50)(50,270,90)
\CArc(150,50)(50,90,270)
\Line(25,100)(275,100)
\Line(25,0)(275,0)
\Text(50,110)[l]{p$_{1}$}
\Text(225,110)[l]{p$_{3}$}
\Text(50,-15)[l]{p$_{2}$}
\Text(225,-15)[l]{p$_{4}$}
\Text(90,50)[l]{q}
\Text(205,50)[l]{p$_{1}$-p$_{3}$-q}
\Text(0,-50)[l]{Fig. 4. $\phi^4$ scattering diagram}
\end{picture}
\end{center}
\end{document}